\documentclass[pre,reprint,amsmath,amssymb,aps,longbibliography]{revtex4-1}
\usepackage{graphicx}
\usepackage{hyperref}
\usepackage{bm}
\usepackage{color}
\usepackage{multirow}
\usepackage{amsfonts}

\newcommand{\tail}{\text{tail}}
\newcommand{\RFA}{\text{RFA}}
\newcommand{\MD}{\text{MD}}
\newcommand{\PY}{\text{PY}}
\newcommand{\num}{\text{num}}
\newcommand{\talpha}{\alpha}
\newcommand{\tomega}{\omega}
\newcommand{\tdelta}{\delta}
\newcommand{\trho}{\rho}
\newcommand{\tA}{A}
\newcommand{\tr}{r}
\newcommand{\tk}{k}
\newcommand{\q}{\sigma}
\newcommand{\MM}{\mu}
\newcommand{\DD}{\mathcal{D}}

\begin{document}
\title{Structural properties of  additive binary hard-sphere mixtures}

\author{S. Pieprzyk}
\email{pieprzyk@ifmpan.poznan.pl}

\author{A. C. Bra\'nka}
\email{branka@ifmpan.poznan.pl}
\affiliation{Institute of Molecular  Physics,
Polish Academy of Sciences, M. Smoluchowskiego 17, 60-179 Pozna\'n, Poland}

\author{S. B. Yuste}
\email{santos@unex.es}

\author{A. Santos}
\email{andres@unex.es}
\affiliation{Departamento de F\'isica and Instituto de Computaci\'on Cient\'ifica Avanzada (ICCAEx), Universidad de Extremadura, Badajoz, E-06006, Spain}

\author{M. L\'opez de Haro}
\email{malopez@unam.mx}
\affiliation{Instituto de Energ\'ias Renovables, Universidad Nacional Aut\'onoma de M\'exico (U.N.A.M.), Temixco, Morelos 62580, Mexico.}

\date{\today}

\begin{abstract}
An approach to obtain the structural properties of  additive binary hard-sphere mixtures is presented. Such an approach, which is a nontrivial generalization of the one recently used for monocomponent hard-sphere fluids [S. Pieprzyk, A. C. Bra\'nka, and D. M. Heyes, Phys.\ Rev.\ E {\bf 95}, 062104 (2017)], combines accurate molecular-dynamics simulation data, the pole structure representation of the total correlation functions, and the Ornstein--Zernike equation. A comparison of the direct correlation functions obtained with the present scheme with those derived  from theoretical results stemming from the Percus--Yevick (PY) closure and the so-called rational-function approximation (RFA) is performed. The density dependence of the leading poles of the Fourier transforms of the total correlation functions and the decay of the pair correlation functions of the mixtures are also addressed and compared to the predictions of the two theoretical approximations.
A very good overall agreement between the results of the present scheme and those of the RFA is found, thus suggesting that the latter (which is an improvement over the PY approximation) can safely be used to
predict reasonably well the long-range behavior, including the structural crossover, of the  correlation functions of  additive binary hard-sphere mixtures.
\\
\end{abstract}
\maketitle
\section{Introduction}
\label{sec1}
Structural properties of the liquid phase are routinely expressed in terms of pair or structural correlation functions (SCFs) \cite{HM06,HM13,BH76}.  An example is the radial distribution function (RDF)  $g(r)$, where $r$ means the interparticle separation. This important structural function can be obtained inter alia from the static structure factor determined through X-ray and neutron scattering experiments \cite{MarchTosiBook1976,HeyesBook1997} and from computer simulations of model liquids.

Because the SCFs are present in many theoretical and experimental treatments, their understanding and mutual relationships are of  great importance. In the case of simple liquids, the SCFs have been a subject of intensive studies for decades, and considerable knowledge has been amassed regarding their structure. In the case of multicomponent liquid mixtures, the studies of structural properties, due to their complexity, are much less developed.

Important practical methods for obtaining SCFs are computer simulations and the liquid-state theories based on the Ornstein--Zernike (OZ) relation \cite{HM06},
\begin{equation}
\label{Eq:OZ}
h(r_{12})=c(r_{12})+\rho \int d \mathbf{r}_3\,h(r_{13})c(r_{23}),
\end{equation}
where $\rho$ is the number density, $c(r)$ is the direct correlation function (DCF), and $h(r) = g(r) - 1 $ is the total correlation function.
The subscripts, $1$, $2$, and $3$ denote the positions of three particles, and  $r_{ij} = |\mathbf{r}_i - \mathbf{r}_j |$ is the separation between particles $i$ and $j$.

A key role for understanding and describing  a liquid structure has been played by various  hard-particle models. Among them, the binary hard-sphere  (BHS) mixture is of particular relevance as it can be considered  as the simplest  model for real liquid mixtures. In  spite of the very simple form of the interparticle potential, the phase diagram of BHS mixtures is fairly complex, their  structural properties being far from trivial and needing further investigations \cite{DRE98,M08}.

In this paper, a framework allowing us to obtain an accurate representation of the SCFs of additive BHS mixtures is proposed. The method includes molecular dynamics (MD) simulation data, residue theorem analysis, and the OZ relation.
The main aim is to obtain a more comprehensive representation of the SCFs for additive BHS, bring to light some new features, and compare results with two analytical predictions. In our approach,  the tail parts of the SCFs are taken into account without using any approximate closures.
We focus here on the DCFs, which are in general quantities very difficult to access from the RDFs due to inherent errors in truncated numerical Fourier transforms \cite{S18}.

Analytical formulas from the Percus--Yevick (PY) approximation for the DCFs of monocomponent hard-spheres and additive BHS mixtures have been available for many decades, and this can be considered as one of the most important results in the history of statistical-mechanical liquid-state theory.
Just one year after Wertheim \cite{W63} and Thiele \cite{T63} found the exact solution of the (three-dimensional) OZ equation with the PY closure for a monocomponent HS fluid, Lebowitz
extended the solution to additive BHS  mixtures \cite{L64}.
In the past two decades, some approximate analytical formulas for the SCFs have been proposed, mainly to reduce limitations of the PY theory.
Yuste et al.\ \cite{YS91,YHS96,YSH98,TH07,HYS08,S16} derived analytic approximations, for both the monocomponent hard-sphere (HS) case and additive HS mixtures, based on a generalization of the PY result.
Their method, usually referred to as  the Rational-Function Approximation (RFA), circumvents the thermodynamic consistency problem of the PY solution.
In fact, the RFA can be seen as an augmented PY solution that includes an extra parameter, $\alpha_{\RFA}$, such that  the PY form is recovered by the choice $\alpha_{\RFA}=0$;
on the other hand, prescribing the isothermal compressibility and the contact values of the RDFs yields a quadratic (monocomponent HS fluid) or a quartic (additive BHS mixture)
equation for $\alpha_{\RFA}$. The RFA method has also been extended to nonadditive HS mixtures \cite{FS11,FS13,FS14}. Here, we implement the RFA approximation with 
the Boubl\'ik--Grundke--Henderson--Lee--Levesque (BGHLL) contact values \cite{B70,GH72,LL73} and
the isothermal compressibility corresponding to
the Boubl\'ik--Mansoori--Carnahan--Starling--Leland (BMCSL) equation of state \cite{B70,MCSL71}.
In this work, those two theoretical approximations (RFA and PY) will be compared with the DCFs obtained from simulation results of the RDFs via our proposed scheme. This in turn will be used to determine the two leading poles of the Fourier transforms of the total correlation functions, and to analyze the structural crossover in additive BHS \cite{ELHH94,SPTER16}.
All of this will allow for an assessment of  the  performance of the RFA as a valuable tool for exploring different density and/or composition regions of BHS mixtures.

The work is organized as follows. In Sec.~\ref{sec2} the general theory of the RDF and DCF is covered, focusing especially on the large-wavenumber limit in Fourier space. The monocomponent case is discussed in Sec.~\ref{sec3A}. In Sec.~\ref{sec3B} the calculation details for additive BHS mixtures are presented and discussed. The main conclusions are summarized in Sec.~\ref{sec4}.

\section{Structural correlation functions}
\label{sec2}
We consider an additive BHS mixture  composed of small ($s$) and big ($b$) particles, characterized by the size ratio $\sigma_s / \sigma_b\leq 1$, where $\sigma_s$ and $\sigma_b$ are HS diameters.
Thus,  there are three different separations between particles at contact: the small-small particle separation, $\sigma_{ss}=\sigma_s$, the small-big particle separation,
$\sigma_{sb}=\frac{1}{2}(\sigma_s +\sigma_b)$, and the big-big particle separation, $\sigma_{bb}=\sigma_b$.
The additive BHS system is defined with the pairwise interaction
\begin{equation}
\label{Eq:potential}
u_{ij}(r) = \left\{
\begin{array}{ll}
\infty, & 0 < r < \sigma_{{ij}}, \\
0,& r > \sigma_{{ij}},
\end{array} \right.
\end{equation}
where $i, j =s, b$.

The partial packing fractions are defined as $\eta_i=\frac{\pi}{6}\rho_i \sigma_i^3$, where $\rho_i=N_i/V$ are number densities, $N_i$ and $V$ being the number of particles of species $i$ and  the volume of the system,  respectively. The total number of particles, number density, and packing fraction are $N=N_{s}+N_{b}$, $\rho = N/V = \rho_s+\rho_b$, and  $\eta=\eta_s+\eta_b$, respectively.

The focus in this work is on an accurate determination of the structural properties  of the additive BHS fluid mixture  by exploiting the pole method \cite{HM13,HM06,ELHH94,GDER04,PBH17}.
From the method it follows inter alia that the asymptotic decay of $h_{ij}(r)$ is determined by the poles of the Fourier transform $\tilde{h}_{ij}(k)$ with the imaginary part closest to the real axis \cite{ELHH94}.
Also, the method was recently shown to be a useful means to obtain the entire DCF and the first two poles in the case of the monocomponent HS fluid \cite{PBH17}. Below, we extend  this  scheme to the additive BHS mixture.

\subsection{Total correlation functions}
The representation of $h_{ij}(r)$ of additive BHS mixtures in terms of the pole structure can be written as \cite{HM06,ELHH94,GDER04,PBH17}
\begin{align}
{h}_{ij}(r) =& -\Theta(\q_{ij}-r)+\Theta(r-\q_{ij})\nonumber\\
&\times\sum_{n=1}^{\infty} {\frac{ \tA_{ij}^{(n)}}{\tr}} e^{-\talpha_{n} \tr } \sin\left(\tomega_{n} \tr +\tdelta_{ij}^{(n)} \right),
\label{Eq:hr_reduced}
\end{align}
where $\Theta(x)$ is the Heaviside step function. The damping coefficients ($\talpha_n$) and the oscillation frequencies ($\tomega_n$) are common to all the pairs, while the amplitudes ($\tA_{ij}^{(n)}$) and the phase shifts ($\tdelta_{ij}^{(n)}$) are specific for each $h_{ij}(r)$ \cite{GDER04}.
It is noteworthy that the Fourier transform,
\begin{equation}
\label{Eq:hkij}
\tilde{h}_{ij}(k)=4 \pi \int_{0}^{\infty} d\tr\,\tr^2 h_{ij}(r) {{\sin(\tk \tr)} \over {\tk \tr}} ,
\end{equation}
of the above representation for $h_{ij}(r)$ can be expressed by the following analytic expression,
\begin{eqnarray}
\label{Eq:hq_ij}
\tilde{h}_{ij}(k) &= & 4 \pi \left[\frac {\q_{ij} \cos( \q_{ij} \tk)}{\tk^2} - \frac{  \sin( \q_{ij} \tk)}{\tk^3}\right] \nonumber\\
&& + {{2 \pi} \over {\tk}} \sum_{n=1}^{\infty}\left[P_{ij}^{(n)}(\q_{ij},k)-P_{ij}^{(n)}(\q_{ij},-k)\right],
\end{eqnarray}
where
\begin{eqnarray}
\label{Eq:Pijn}
P_{ij}^{(n)}(\q_{ij},k)&\equiv&  \frac{\tA_{ij}^{(n)} e^{-\talpha_{n}\q_{ij} }}{\talpha_{n}^2 +\left(\tomega_{n}-\tk  \right)^2  }\left\{ \talpha_{n}\cos[\tdelta_{ij}^{(n)}+(\tomega_{n}-\tk)\q_{ij}]\right.\nonumber\\
&&\left.-(\tomega_{n}-\tk ) \sin[\tdelta_{ij}^{(n)}+(\tomega_{n}-\tk )\q_{ij}]\right\}.
\end{eqnarray}

\subsubsection{Large-${k}$ limit }

The large-${k}$ limit or `tail' of  $\tilde{h}_{ij}({k})$ can be obtained from  Eq.~(\ref{Eq:hq_ij})  by expanding in negative powers of $\tk$
the functions multiplying the trigonometric functions $\sin(\q_{ij}\tk)$ and $\cos(\q_{ij}\tk)$, namely
\begin{equation}
\label{Eq:hg_tailij}
\tilde{h}_{ij}^{\tail}(k)  = \cos(\q_{ij} \tk ) \sum_{n=1}^{\infty} {{C_{ij}^{(n)}}\over{ \tk ^{2n}}}
+\sin(\q_{ij} \tk ) \sum_{n=1}^{\infty} {{D_{ij}^{(n)}}\over{ \tk ^{2n+1}}}.
\end{equation}
The first few coefficients $C_{ij}^{(n)}$ and $D_{ij}^{(n)}$  are given in
Appendix \ref{appA}, where it is also shown that those coefficients can be expressed in terms of derivatives of the RDF at contact (i.e., at $\tr=\q_{ij}^+$) as
\begin{subequations}
\label{Eq:hg_tail_C1-D2}
\begin{eqnarray}
\label{Eq:hg_tail_C1}
C_{ij}^{(1)} &=& 4 \pi \q_{ij} g_{ij}(\q_{ij}^+), \\
\label{Eq:hg_tail_D1}
D_{ij}^{(1)} &=& -4\pi \left[ g_{ij}(\q_{ij}^+) + \q_{ij} {g}'_{ij}(\q_{ij}^+) \right], \\
\label{Eq:hg_tail_C2}
C_{ij}^{(2)} &=& -4\pi \left[ 2 {g}'_{ij}(\q_{ij}^+) + \q_{ij} g_{ij}''(\q_{ij}^+)\right], \\
\label{Eq:hg_tail_D2}
D_{ij}^{(2)} &=& 4\pi \left[ 3 g_{ij}''(\q_{ij}^+) + \q_{ij} g_{ij}'''(\q_{ij}^+) \right],
\end{eqnarray}
\end{subequations}
where single, double, and triple primes represent the first, second, and third derivatives of the RDF, respectively.

\subsubsection{Small-$k$ limit}

It can be shown from Eq.\ \eqref{Eq:hq_ij} that $\tilde{h}_{ij}(k)$ is an even function regular at $k=0$, so that its Taylor expansion in powers of $k$  is
\begin{equation}
\tilde{h}_{ij}(k) = \tilde{h}_{ij}^{(0)} + \tilde{h}_{ij}^{(2)} \tk ^2 + \tilde{h}_{ij}^{(4)} \tk ^4 + \cdots ,
\label{Eq:hq_small}
\end{equation}
where the zeroth-order term is
  \begin{align}
\label{Eq:hq_small_h0}
\tilde{h}_{ij}^{(0)}=& -{{4 \pi}\over {3}}\q_{ij}^3 + 4 \pi \sum_{n=1}^{\infty} {{\tA_{ij}^{(n)} e^{-\talpha_{n}\q_{ij}}}\over{(\talpha_{n}^2+\tomega_{n}^2)^2}} \nonumber\\
 &\times\left[\tomega_{n}( 2\talpha_{n}+ \talpha_{n}^2  \q_{ij} + \tomega_{n}^2 \q_{ij}) \cos(\tdelta_{ij}^{(n)}+\tomega_{n}\q_{ij})
\right.\nonumber \\
& \left.+ (\talpha_{n}^2+ \talpha_{n}^3 \q_{ij}- \tomega_{n}^2 + \talpha_{n} \tomega_{n}^2 \q_{ij}) \sin(\tdelta_{ij}^{(n)}+\tomega_{n}\q_{ij}) \right].
\end{align}
The coefficients of higher order are similar to the zeroth-order one but have more complex expressions.
The coefficients $\tilde{h}_{ij}^{(0)}$ for all $ij$ are involved in the isothermal compressibility $\kappa_T$ \cite{S16}.

\subsection{Direct correlation functions}

The OZ relation, Eq.\ (\ref{Eq:OZ}), for binary mixtures in Fourier space, has the form \cite{S16}
\begin{equation}
\label{Eq:OZ1}
\tilde{h}_{ij}(k)  = \tilde{c}_{ij}(k) +  \sum_\ell \trho_\ell \tilde{c}_{i\ell}({k}) \tilde{h}_{\ell j}({k}),
\end{equation}
and in matrix notation,
\begin{equation}
\label{Eq:OZ4}
\widehat{\mathsf{h}}(\tk) = \widehat{\mathsf{c}}(\tk)\cdot\left[\mathsf{I}+\widehat{\mathsf{h}}(\tk)\right],
\end{equation}
where
$\widehat{h}_{ij}(\tk)\equiv \sqrt{\trho_{i}\trho_{j}}\tilde{h}_{ij}(\tk)$, $\widehat{c}_{ij}(\tk)\equiv \sqrt{\trho_{i}\trho_{j}}\tilde{c}_{ij}(\tk)$, and $\mathsf{I}$ is the identity matrix.
Thus,
\begin{equation}
\label{Eq:cq1}
 \widehat{\sf c}(\tk)= \widehat{\sf h}(\tk)\cdot[{\sf
I}+\widehat{\sf h}(\tk)]^{-1}={\sf
I}-[{\sf
I}+\widehat{\sf h}(\tk)]^{-1}.
\end{equation}
More explicitly, one has
\begin{subequations}
\label{Eq:cq11-Eq:cq12}
\begin{equation}
\label{Eq:cq11}
\tilde{c}_{ss}({k})={{\tilde{h}_{ss}({k}) +\trho_b\left[\tilde{h}_{ss}({k}) \tilde{h}_{bb}({k})- \tilde{h}^2_{sb}({k})\right]} \over {\DD({k})}},
\end{equation}
\begin{equation}
\label{Eq:cq12}
\tilde{c}_{sb}({k})={{\tilde{h}_{sb}({k})} \over {\DD({k})}},
\end{equation}
\end{subequations}
where
\begin{align}
\DD({k})=&1 + \trho_s \tilde{h}_{ss}({k}) + \trho_b \tilde{h}_{bb}({k})  \nonumber\\
&+ \trho_s \trho_b \left[\tilde{h}_{ss}({k}) \tilde{h}_{bb}({k})-\tilde{h}_{sb}^2({k})\right].
\end{align}
For the sake of conciseness, we omit the expression for $\tilde{c}_{bb}({k})$, which can be obtained from Eq.\ \eqref{Eq:cq11} by the simple exchange $s\leftrightarrow b$. Henceforth, we will do the same for any quantity of the form $X_{bb}$, which can then be obtained from $X_{ss}$ by the same exchange of indices.

\subsubsection{Large-${k}$ limit }
Taking into account that the absolute value of $\DD^\tail\equiv \DD-1$ is smaller than $1$,  we can expand $\DD^{-1}=1-\DD^\tail+{\DD^{\tail}}^2+\cdots$ for large wave number $k$
and use the tail forms as in Eq.\ \eqref{Eq:hg_tailij}. As a consequence, the first few terms in
the large-$k$ limit ("tail") of $\tilde{c}_{ij}(k)$ are
\begin{widetext}
\begin{subequations}
\label{Eq:cqtail11rozw-Eq:cqtail21rozw}
\begin{eqnarray}
\label{Eq:cqtail11rozw}
\tilde{c}^{\tail}_{ss}({k})&=&
\frac{C_{ss}^{(1)}}{\tk ^2} \cos(\q_{s} \tk )  + \frac{D_{ss}^{(1)}}{\tk ^3}  \sin(\q_{s} \tk ) +
\frac{C_{ss}^{(2)}}{\tk ^4}  \cos(\q_{s} \tk ) - \trho_s  \frac{{C_{ss}^{(1)}}^2}{\tk ^4}  \cos^2(\q_{s} \tk ) - \trho_b  \frac{{C_{sb}^{(1)}}^2}{\tk ^4}  \cos^2(\q_{sb} \tk ) \nonumber \\
&&+
\frac{D_{ss}^{(2)}}{\tk ^5}  \sin(\q_{s} \tk ) - 2\trho_s  \frac{ C_{ss}^{(1)} D_{ss}^{(1)}}{\tk ^5}  \sin(\q_{s} \tk ) \cos(\q_{s} \tk ) - 2\trho_b  \frac{ C_{sb}^{(1)} D_{sb}^{(1)}}{\tk ^5}  \sin(\q_{sb} \tk ) \cos(\q_{sb} \tk ) + \cdots,
\end{eqnarray}
\begin{eqnarray}
\label{Eq:cqtail21rozw}
\tilde{c}^{\tail}_{sb}({k})&=&
\textcolor{black}{ {{C_{sb}^{(1)}} \over {\tk ^2}} \cos(\q_{sb} \tk ) } + {{D_{sb}^{(1)}} \over {\tk ^3}}  \sin(\q_{sb} \tk )
+{{C_{sb}^{(2)}} \over {\tk ^4}}  \cos(\q_{sb} \tk ) -    {{C_{sb}^{(1)} } \over {\tk ^4}}  \cos(\q_{sb} \tk )\left[\trho_s C_{ss}^{(1)}\cos(\q_{s} \tk )+\trho_b C_{bb}^{(1)}\cos(\q_{b} \tk )\right]   \nonumber \\
&& + {{D_{sb}^{(2)}} \over {\tk ^5}}  \sin(\q_{sb} \tk ) -   {{D_{sb}^{(1)}  } \over {\tk ^5}}  \sin(\q_{sb} \tk ) \left[ \trho_s C_{ss}^{(1)}\cos(\q_{s} \tk )+\trho_b C_{bb}^{(1)}\cos(\q_{b} \tk )\right]  \nonumber \\
&&-   {{ C_{sb}^{(1)} } \over {\tk ^5}}   \cos(\q_{sb} \tk ) \left[\trho_s D_{ss}^{(1)}\sin(\q_{s} \tk )+\trho_b D_{bb}^{(1)}\sin(\q_{b} \tk )\right]+\cdots.
\end{eqnarray}
\end{subequations}
\end{widetext}

\subsubsection{Evaluation of the function $c_{ij}({r})$}
From the Fourier transform $\tilde{c}_{ij}(\tk)$ one can obtain the DCFs in real space as
\begin{equation}
c_{ij}({r})={1 \over {2 \pi^2}} \int_{0}^{\infty} d\tk\,\tk ^2 \tilde{c}_{ij}({k}) {{\sin{(\tk \tr)}} \over {\tk \tr}}.
\end{equation}
At a practical level, it is useful to introduce an arbitrarily large  wave number $Q$ and decompose $c_{ij}({r})$ into two contributions, namely
\begin{equation}
 c_{ij}({r})= c^{\num}_{ij}({r}) + c^{\tail}_{ij}({r}),
\end{equation}
where
\begin{subequations}
 \begin{equation}
\label{Eq:crN}
c^{\num}_{ij}({r})={1 \over {2 \pi^2}} \int_{0}^{Q} d\tk\,\tk ^2 \tilde{c}_{ij}({k}) {{\sin{(\tk \tr)}} \over {\tk \tr}}  ,
\end{equation}
\begin{equation}
\label{Eq:crA}
c^{\tail}_{ij}({r})={1 \over {2 \pi^2}} \int_{Q}^{\infty}d\tk\, \tk ^2 \tilde{c}^{\tail}_{ij}({k}) {{\sin{(\tk \tr)}} \over {\tk \tr}}  .
\end{equation}
\end{subequations}
The first contribution, Eq.\ \eqref{Eq:crN}, can be obtained numerically (we have used the five-point method of integration) from $\tilde{c}_{ij}({k})$ given in Eqs.\ \eqref{Eq:cq11-Eq:cq12}. In contrast, the second contribution, Eq.\ \eqref{Eq:crA}, can be evaluated analytically term by term [see Appendix \ref{appB0}, where the first few terms contributing to $c_{ij}^\tail(r)$ are explicitly given].

Let us stress that the tail functions $c^{\tail}_{ij}({r})$   [and, consequently, the functions $\tilde{c}^{\tail}_{ij}({k})$ or, equivalently, $\tilde{h}^{\tail}_{ij}({k})$] contain relevant information that cannot be omitted in any accurate representation of the DCFs. 

\subsubsection{Discontinuities at $r=\sigma_{ij}$}
As shown in Appendix \ref{appB0}, $c^{\tail}_{ij}({r})$ presents a zeroth-order singularity  (jump) at $r=\q_{ij}$. Since those discontinuities are independent of $Q$, and $c^{\num}_{ij}({r})$ is continuous, it turns out that the discontinuities of $c^{\tail}_{ij}({r})$ determine those of the full functions $c_{ij}({r})$. From the results of Appendix \ref{appB0}, one gets
\begin{subequations}
\label{disc_cij_contact}
\begin{align}
\Delta c_{ij}(\q_{ij})=&\frac{C_{ij}^{(1)}}{4\pi\q_{ij}}
=g_{ij}(\q_{ij}^+),\\
\Delta c_{ij}'(\q_{ij})=&-\frac{1}{4\pi}\left(\frac{C_{ij}^{(1)}}{\q_{ij}^2}+\frac{D_{ij}^{(1)}}{\q_{ij}}\right)
=g_{ij}'(\q_{ij}^+),\\
\Delta c_{ij}''(\q_{ij})=&\frac{1}{4\pi}\left(\frac{2C_{ij}^{(1)}}{\q_{ij}^3}+\frac{2D_{ij}^{(1)}}{\q_{ij}^2}-\frac{C_{ij}^{(2)}}{\q_{ij}}\right)=g_{ij}''(\q_{ij}^+),\\
\Delta c_{ij}'''(\q_{ij})=&\frac{1}{4\pi}\left(
-\frac{6C_{ij}^{(1)}}{\q_{ij}^4}-\frac{6D_{ij}^{(1)}}{\q_{ij}^3}+\frac{3C_{ij}^{(2)}}{\q_{ij}^2}
+\frac{D_{ij}^{(2)}}{\q_{ij}}
\right)
\nonumber\\
=&g_{ij}'''(\q_{ij}^+),
\end{align}
\end{subequations}
where we have introduced the short-hand notation $\Delta X(a)\equiv \lim_{r\to a^+}X(r)-\lim_{r\to a^-}X(r)$ and in the last steps use has been made of Eqs.\ \eqref{Eq:hg_tail_C1-D2}. Equations \eqref{disc_cij_contact} are consistent with the continuity of the indirect correlation functions $h_{ij}(r)-c_{ij}(r)$  and their first three derivatives at $r=\q_{ij}$.

\begin{figure}
	\includegraphics[width=0.9\columnwidth]{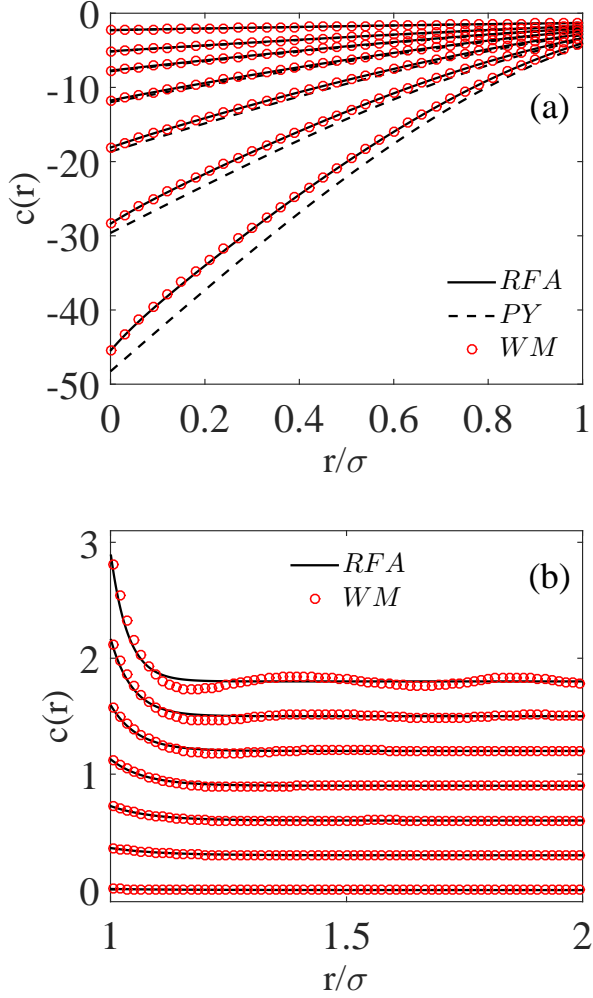}
	\caption{DCF for the monocomponent  HS fluid in the regions (a) $0 <r/\sigma< 1$ and (b) $1 < r/\sigma < 2$. The densities are, from top to bottom in (a) and from bottom to top in (b), $\rho\sigma^3 = 0.2, 0.4, 0.5, 0.6, 0.7, 0.8, 0.9$. In panel (b), the curves have been  shifted vertically for better clarity. The open circles represent the results obtained from the WM-scheme (which combines MD data with the pole representation as described in Sec. \ref{sec3}), the dashed lines are from the PY approximation, and  the solid lines are from the RFA. Note that $c^\PY(r)=0$ for $r/\q>1$.}
	\label{fig1}
\end{figure}

\subsection{Determination of the poles of $\tilde{h}_{ij}(k)$}
{}From the OZ relation \eqref{Eq:OZ4} it is straightforward to obtain $\widehat{\mathsf{h}}(k)=\left[\mathsf{I}-\widehat{\mathsf{c}}(k)\right]^{-1}\cdot \widehat{\mathsf{c}}(k)$. Therefore, the poles $\tk = \pm \tomega + \imath \talpha$ of $\tilde{h}_{ij}(k)$ are given by the zeros of the determinant of $\mathsf{I}-\widehat{\mathsf{c}}(k)$, i.e., $ D(k)\equiv[1-\trho_s \tilde{c}_{ss}({k})][1-\trho_b \tilde{c}_{bb}({k})] - \trho_s \trho_b \tilde{c}_{sb}^2({k})=0$.
By equating real and imaginary parts, the formulas
\begin{subequations}
\label{Eq:poles1-Eq:poles2}
\begin{equation}
\label{Eq:poles1}
1=\trho_s {I}_{s}^{(0)}(\alpha,\omega) + \trho_b {I}_{b}^{(0)}(\alpha,\omega) + \trho_s \trho_b  {I}_{sb}^{(0)}(\alpha,\omega),
\end{equation}
\begin{equation}
\label{Eq:poles2}
1=\trho_s {I}_{s}^{(1)}(\alpha,\omega) + \trho_b {I}_{b}^{(1)}(\alpha,\omega) + \trho_s \trho_b  {I}_{sb}^{(1)}(\alpha,\omega),
\end{equation}
\end{subequations}
are obtained. The required integrals $I_{i}^{(n)}$ and $I_{sb}^{(n)}$ are presented in Appendix \ref{appB}.
If the DCFs, $c_{ij}({r})$, are known and decay sufficiently fast, the above equations provide a practical route to obtain the poles of $\tilde{h}_{ij}(k)$.

\subsection{Determination of amplitudes and phases}
\label{sec2d}
To determine $h_{ij}(r)$ in Eq.\ \eqref{Eq:hr_reduced}, the amplitudes, $A_{ij}^{(n)}$, and phase, $\delta_{ij}^{(n)}$,  are required as well.
The appropriate prescription for their evaluation  was constructed by Evans et al.\ \cite{ELHH94} by application of the residue theorem. In the case of a binary mixture,  the contribution to $rh_{ij}(r)$ associated with the pair of poles $k_{\text{pole}}=\tomega + \imath \talpha$ and $-k_{\text{pole}}^*=-\tomega + \imath \talpha$ is $\bar{A}_{ij}e^{\imath k_{\text{pole}} r}+\bar{A}_{ij}^*e^{-\imath k_{\text{pole}}^* r}$, the `complex amplitudes' $\bar{A}_{ij}$ being given by the following expressions:
\begin{equation}
\label{Eq:ampltude}
\bar{A}_{ss} = \frac{k_{\text{pole}}\left(1-\rho_b \bar{c}_{bb}\right)}{2\pi \rho_s \bar{D}'},\quad \bar{A}_{sb} = \frac{k_{\text{pole}} \bar{c}_{sb}}{2\pi \bar{D}'},
\end{equation}
where
\begin{subequations}
\begin{equation}
\bar{D}'= -\rho_s \left(1- \rho_b \bar{c}_{bb}\right) \bar{c}'_{ss} - \rho_b \left(1 - \trho_s \bar{c}_{ss}\right) \bar{c}'_{bb} - 2 \rho_s \rho_b \bar{c}_{sb} \bar{c}'_{sb},
\end{equation}
\begin{eqnarray}
\bar{c}_{ij}&=&\frac{4 \pi}{k_{\text{pole}}} \int_0^{\infty} dr\, r c_{ij}(r) \left[\cosh(\talpha r) \sin(\tomega r) \right. \nonumber \\
&& \left. + \imath  \sinh(\talpha r) \cos(\tomega r) \right],
\end{eqnarray}
\begin{eqnarray}
\bar{c}_{ij}'&=&-\frac{\bar{c}_{ij}}{k_{\text{pole}}}+\frac{4\pi}{k_{\text{pole}}} \int_0^{\infty} dr\, r^2 c_{ij}(r)\left[ \cosh(\talpha r) \cos(\tomega r) \right. \nonumber \\
&& \left. - \imath  \sinh(\talpha r) \sin(\tomega r)\right].
\end{eqnarray}
\end{subequations}
Note that $\bar{c}_{ij}=\tilde{c}_{ij}(k_{\text{pole}})$, $\bar{c}'_{ij}=\tilde{c}_{ij}'(k_{\text{pole}})$, and $\bar{D}'=D'(k_{\text{pole}})$.
Provided the DCFs are known and the poles are determined from Eqs.\ \eqref{Eq:poles1-Eq:poles2},  then the amplitudes and phases can be evaluated from the real and imaginary parts of the complex amplitudes $\bar{A}_{ij}=|\bar{A}_{ij}|e^{\imath(\delta_{ij}-\frac{\pi}{2})}$ [so, one obtains e.g., $A_{ij}=2|\bar{A}_{ij}|$]. In this way, at least in principle, a contribution of each $n$-pole defined by $\left\lbrace A_{ij}^{(n)}, \talpha_n, \tomega_n, \delta_{ij}^{(n)} \right\rbrace$  to the $h_{ij}(r)$ may be  determined.
In practice, the scheme is strongly limited by the accuracy of the DCFs, and even obtaining the contribution of  the leading poles becomes a hard task. In this work, it is done for the two leading poles of the additive BHS mixture.

\begin{figure}
\includegraphics[width=0.9\columnwidth]{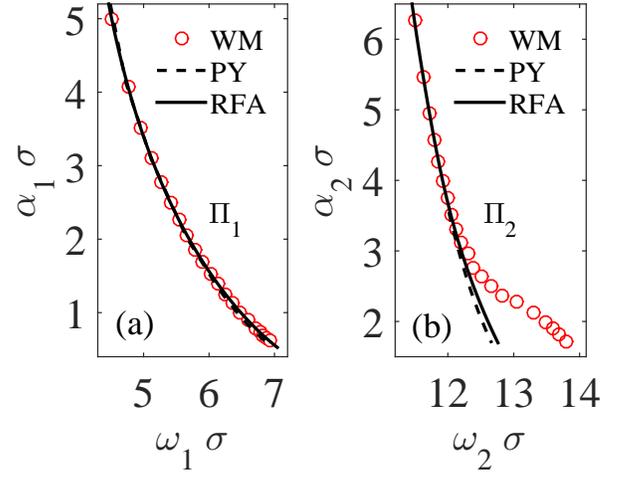}
\caption{(a) First ($\Pi_1$) and (b) second ($\Pi_2$) poles of the monocomponent HS system for a number of different densities representative of the entire fluid region, from  $\rho\sigma^3 = 0.05$ (top point in each panel) to $\rho\sigma^3 = 0.94$ (bottom point  in each panel). 
Here, $\alpha_n$ and $\omega_n$ denote the damping coefficients and the oscillation frequencies, respectively (see Eq.\ \eqref{Eq:hr_reduced}). The open circles are the results obtained  from the WM-scheme in Ref.\ \cite{PBH17} (i.e., the solution of Eqs.\ \eqref{Eq:poles1-Eq:poles2} for the monocomponent HS fluid). The dashed lines are from the PY approximation, and  the solid lines are from the RFA.}
\label{fig2}
\end{figure}

\section{Results} 
\label{sec3}

In this Section we provide the analysis of
the RDF and DCF, with special emphasis on the large-wave-number limit in Fourier space. First we deal with the monocomponent case and subsequently we consider additive BHS mixtures.

\subsection{Monocomponent HS fluid}
\label{sec3A}
The monocomponent case was considered in Ref.\ \onlinecite{PBH17} and the associated SCFs were determined with the so-called WM-scheme (see Sec.\ \ref{sec3B}), which combines the OZ equation, the residue theorem analysis, and simulation data. Here, for the sake of completeness, we revisit this monocomponent case.
Figure \ref{fig1} shows the DCF at several densities as obtained from the WM-scheme and as given by the PY and RFA theoretical approaches. While the PY approximation predicts well the low-density behavior of the HS fluid, its agreement with the WM-determined DCF deteriorates on increasing density.
This  known feature of the PY solution is considerably corrected by the RFA.
As seen in Fig.\ \ref{fig1}(a), in the core region  $ 0 < r/\sigma < 1$  (where $\sigma$ is the diameter of the spheres) the DCF points calculated with the WM-scheme  follow practically exactly  the prediction of the RFA. It is noteworthy that
this excellent agreement takes place for all fluid densities, including those close to the freezing region.

The performance of the RFA at larger $r$-separations is also very good, as observed from  Fig.\ \ref{fig1}(b). On the other hand, small deviations between the monotonic (RFA) and the oscillatory (WM-scheme) results occur, although they are visible only for high densities.

In Fig.\ \ref{fig2}, the density dependences of the first ($\Pi_1$) and second ($\Pi_2$) poles  determined with the WM-scheme  are compared  with those obtained from the PY and RFA approaches.
The RFA reproduces almost perfectly the density dependence of the first pole, and in this respect it  improves upon the outcome of the PY approximation.
The results for the second pole [see Fig.\ \ref{fig2}(b)] demonstrate the non-negligible role of  the DCF part for $r/\sigma > 1$ at  sufficiently high densities. More precisely, the departure of the WM-scheme values and the theoretical ones occurs for $\rho\sigma^3 > 0.7$.
It is worth mentioning that Statt et al.'s analysis for a one-component colloidal suspension yielded a plot similar to Fig.\ \ref{fig2}(a) with simulation and experiment deviating from PY theory at very high packings \cite{SPTER16}.

\subsection{Additive binary hard-sphere mixtures}
\label{sec3B}
In order to obtain accurate DCFs of additive BHS mixtures, the following analytic representation of $h_{ij}({r})$ in the form of two functional parts is considered,
\begin{equation}
\label{Eq:hrWM}
h^{WM}_{ij}({r}) = \left\{
\begin{array}{ll}
-1, & 0 < \tr < \q_{ij}, \\
h^W_{ij}({r}),&\q_{ij} < \tr < \tr_{ij}^{{\min}}, \\
h^M_{ij}({r}),& r>\tr_{ij}^{{\min}} ,
\end{array} \right.
\end{equation}
where
\begin{subequations}
\label{Eq:hrW&M}
  \begin{equation}
\label{Eq:hrW}
h^W_{ij}({r})=\sum\limits_{n=1}^{W} b_{ij}^{(n)} \tr^{n-1},
\end{equation}
\begin{equation}
\label{Eq:hrM}
h^M_{ij}({r})=\sum\limits_{n=1}^{M} {\frac {\tA_{ij}^{(n)}}{\tr}} e^{-\talpha_{n} \tr} \sin(\tomega_{n} \tr + \tdelta_{ij}^{(n)}).
\end{equation}
\end{subequations}
As explained below, the parameters $\{ b_{ij}^{(1)}, b_{ij}^{(2)}, \ldots, b_{ij}^{(W)} \}$ and $\{ \tA_{ij}^{(1)}, \talpha_{1}, \tomega_{1}, \tdelta_{ij}^{(1)}, \ldots, \tA_{ij}^{(M)}, \talpha_{M}, \tomega_{M},\tdelta_{ij}^{(M)} \}$ are obtained by a fitting procedure.

The form of  $h^W_{ij}(r)$ in Eq.~(\ref{Eq:hrW}) is fairly arbitrary, but we seek a rather simple function which provides sufficient flexibility at the next stages of the calculation. In this respect, the polynomial form, the Fourier transform of which can be obtained analytically, is a convenient and appropriate form.
A suitable choice for $\tr_{ij}^{{\min}}$ is the position of the first minimum of $h_{ij}({r})$.
Also, our tests suggest that, for most studied densities, the optimal choices for  $W$ and $M$ are in the range of, approximately, $8$--$15$ and in this way the final results are fairly insensitive to the particular values of those parameters. In the results presented below we have usually taken $W=15$ and $M=10$.

Furthermore, the function $h_{ij}^{WM}$ and its first derivative are constrained to be continuous at $r=\tr_{ij}^{{\min}}$, so that the following continuity conditions are imposed in the scheme,
\begin{subequations}
\begin{equation}
\label{Eq:hrWM_conditions1}
h^W_{ij}(\tr_{ij}^{{\min}})=h^M_{ij}(\tr_{ij}^{{\min}}),
\end{equation}
\begin{equation}
\label{Eq:hrWM_conditions2}
\left. {{\partial h^W_{ij}({r})} \over {\partial r}} \right|_{\tr_{ij}^{{\min}}}=\left. {{\partial h^M_{ij}({r})} \over {\partial r}} \right|_{\tr_{ij}^{{\min}}}=0.
\end{equation}
\end{subequations}
Moreover, the contact values proposed, independently, by Boubl\'ik \cite{B70}, Grundke and Henderson \cite{GH72}, and Lee and Levesque \cite{LL73} are enforced, so that
\begin{eqnarray}
\label{Eq:hrWM_conditions3b}
h_{ij}^W(\q_{ij}^+)&=&\frac{\eta}{1-\eta} + \frac{3}{2} \frac{\eta}{(1-\eta)^2} \frac{\sigma_i \sigma_j}{\sigma_{ij}} \frac{\MM_2}{\MM_3}
\nonumber \\
&& +\frac{1}{2} \frac{\eta^2}{(1-\eta)^3} \left( \frac{\sigma_i \sigma_j}{\sigma_{ij}} \frac{\MM_2}{\MM_3} \right)^2,\quad \MM_n\equiv \frac{1}{\rho}\sum_i \rho_i\sigma_i^n.\nonumber\\
\end{eqnarray}

The Fourier transform of the above representation of $h^{WM}_{ij}({r})$ in Eq.\ (\ref{Eq:hrWM}) is given by the analytic expression
\begin{eqnarray}
\label{Eq:hqWM}
\tilde{h}^{WM}_{ij}({k}) &=& 4 \pi \left[{{ \q_{ij} \cos(\tk  \q_{ij})} \over {\tk ^{2}}} - {{\sin(\tk  \q_{ij})} \over {\tk ^3}} \right]+\tilde{h}^{W}_{ij}({k})
\nonumber\\
&&+\tilde{h}^{M}_{ij}({k}),
\end{eqnarray}
where
\begin{subequations}
\label{Eq:hqW&M}
\begin{eqnarray}
\label{Eq:hqW}
\tilde{h}^{W}_{ij}({k}) &=&\sum_{n=1}^{[\frac{W+3}{2}]}\frac{4\pi}{k^{2n}}\Big[s_{ij}^{(n)}(\q_{ij})\cos(k\q_{ij})-s_{ij}^{(n)}(r_{ij}^{\min})
\nonumber\\
&&\times \cos(kr_{ij}^{\min})\Big]
+\sum_{n=1}^{[\frac{W+2}{2}]}\frac{4\pi}{k^{2n+1}}\Big[t_{ij}^{(n)}(\q_{ij})\sin(k\q_{ij})\nonumber\\
&&-t_{ij}^{(n)}(r_{ij}^{\min})\sin(kr_{ij}^{\min})\Big],
\end{eqnarray}
\begin{equation}
\label{Eq:hqM}
\tilde{h}^{M}_{ij}({k}) = \frac{2 \pi}{\tk } \sum_{n=1}^{M}\left[ P_{ij}^{(n)}(\tr_{ij}^{{\min}},k)-P_{ij}^{(n)}(\tr_{ij}^{{\min}},-k)\right].
\end{equation}
\end{subequations}
In the upper summation limits of Eq.\ \eqref{Eq:hqW}, $[\cdots]$ denotes the integer part, and the coefficients $s_{ij}^{(n)}(a)$ and $t_{ij}^{(n)}(a)$ (with $a=\q_{ij}$ and $a=r_{ij}^{\min}$) are linear combinations of the coefficients $b_{ij}^{(n)}$ whose explicit expressions will be omitted here for the sake of simplicity. We recall that in Eq.\ \eqref{Eq:hqM} $P_{ij}^{(n)}$ is defined by Eq.\ \eqref{Eq:Pijn}.

\begin{figure*}
\includegraphics[width=0.66\columnwidth]{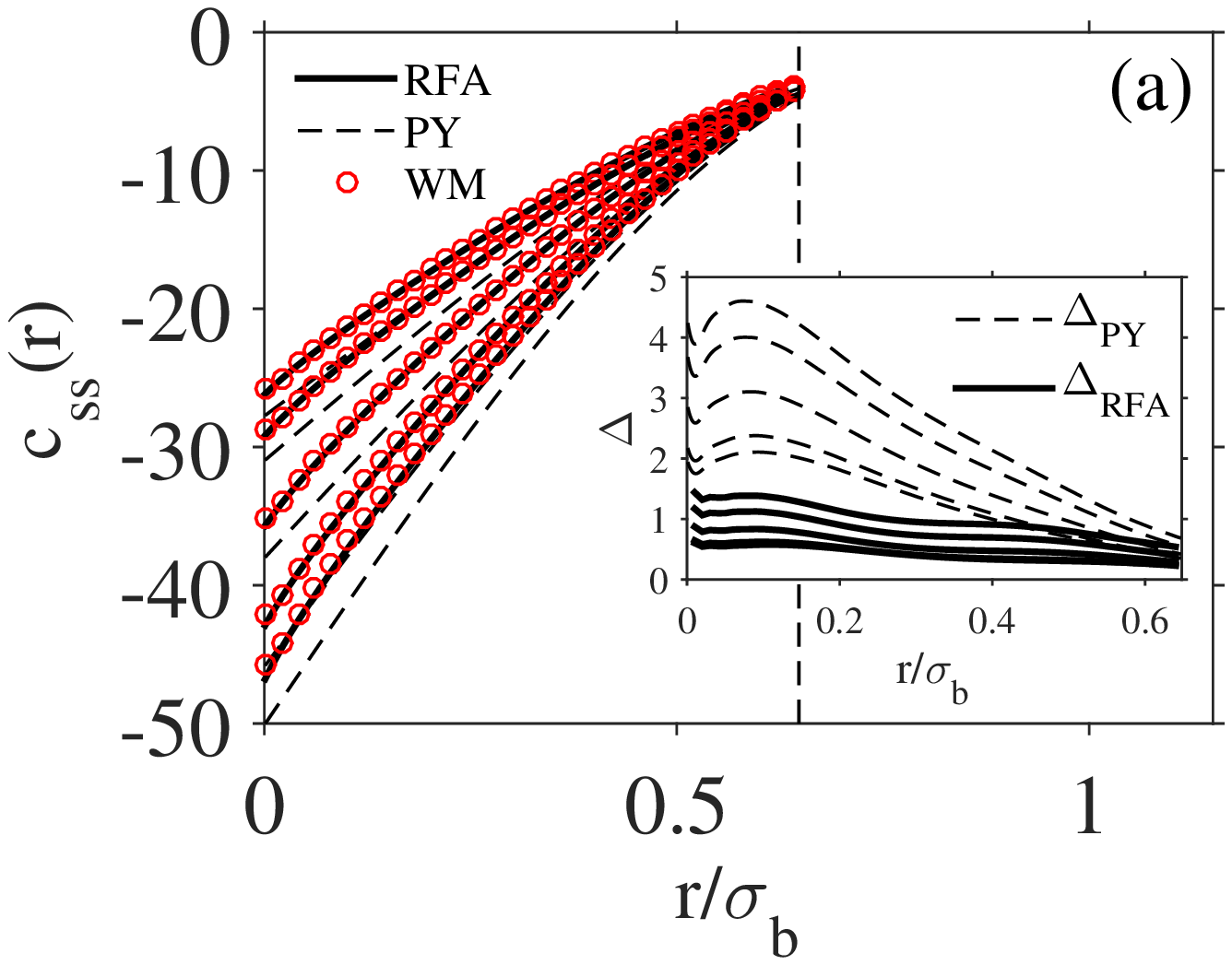}
\includegraphics[width=0.66\columnwidth]{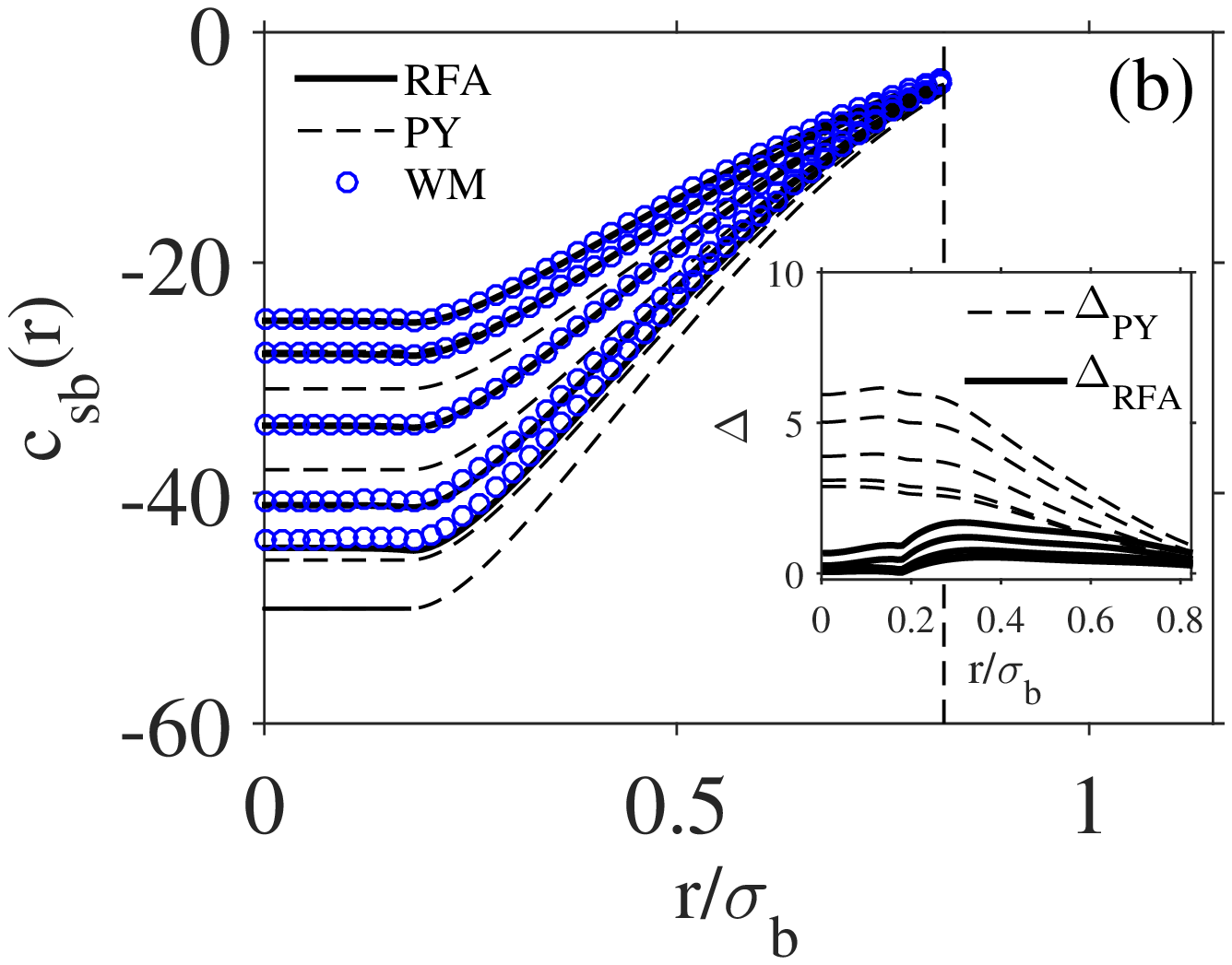}
\includegraphics[width=0.66\columnwidth]{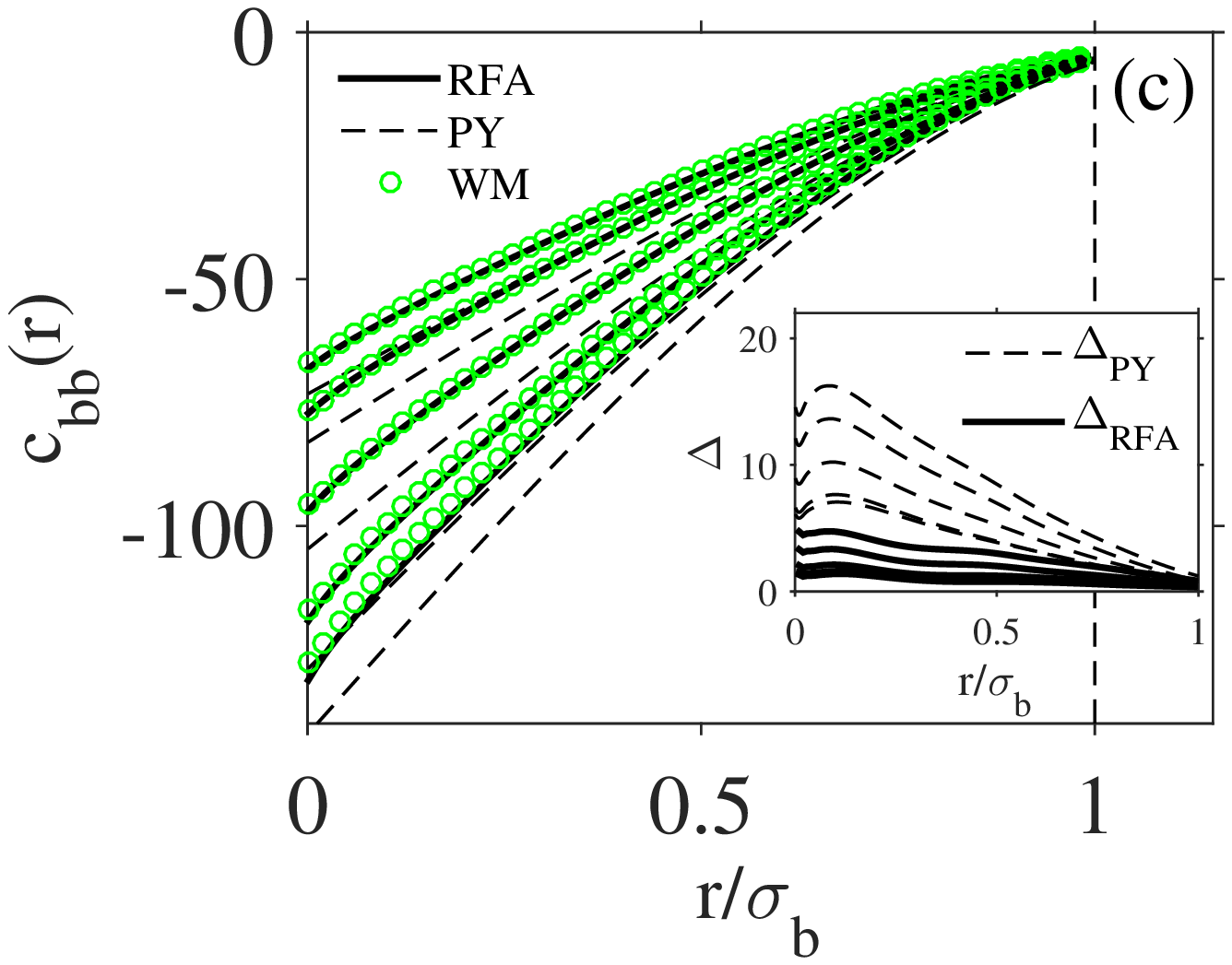}
\caption{Plots of (a) small-small, (b) small-big, and (c) big-big DCFs in the region $0 < r < \q_{ij}$ for additive BHS mixtures with a size ratio $\q_s/\q_b=0.648$ and a  total packing fraction $\eta=0.5$. The partial packing fractions are, from top to bottom, $\eta_s = 0.05, 0.10, 0.20, 0.30, 0.35$.
The open  circles represent the results obtained from the WM-scheme, the dashed lines are from the PY approximation, and the solid  lines are from the RFA.
In the insets, the differences $ \Delta_{\PY} = c^{WM}_{ij}(r) - c^{\PY}_{ij}(r)$ and $ \Delta_{\RFA} = c^{WM}_{ij}(r) - c^{\RFA}_{ij}(r)$ are shown, those differences increasing with increasing $\eta_s$.}
	\label{fig3}
\end{figure*}

\begin{figure*}
\includegraphics[width=0.66\columnwidth]{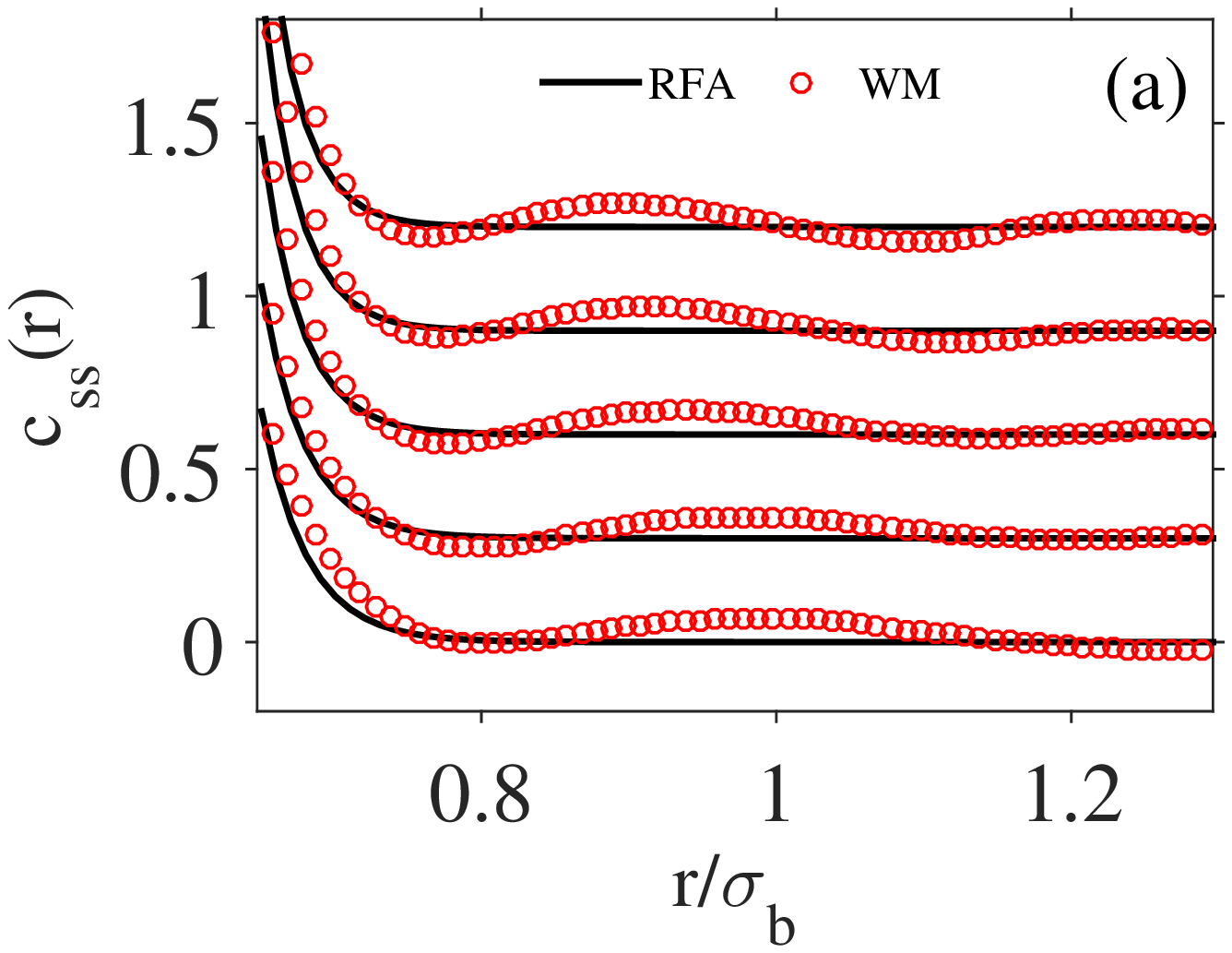}
\includegraphics[width=0.66\columnwidth]{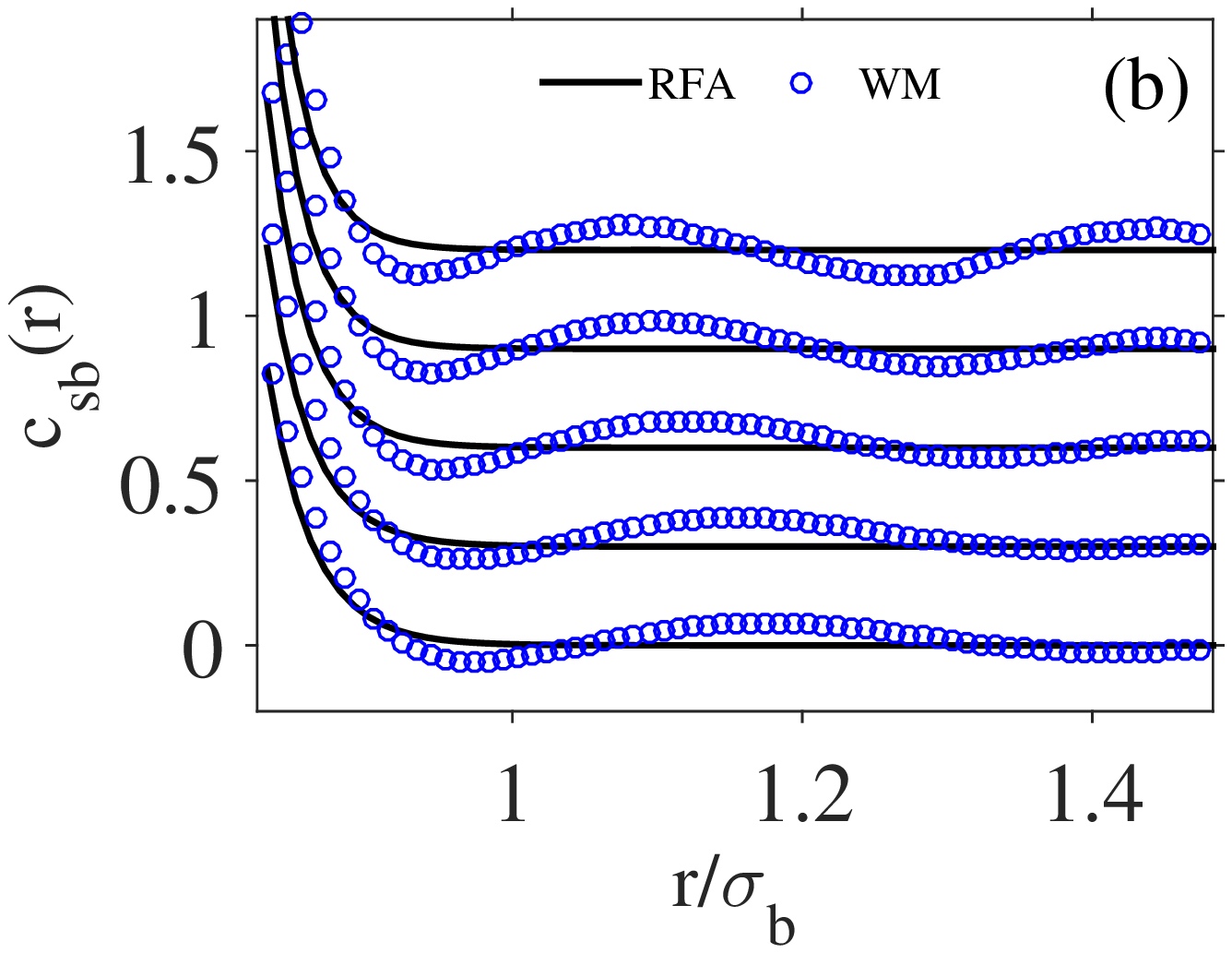}
\includegraphics[width=0.66\columnwidth]{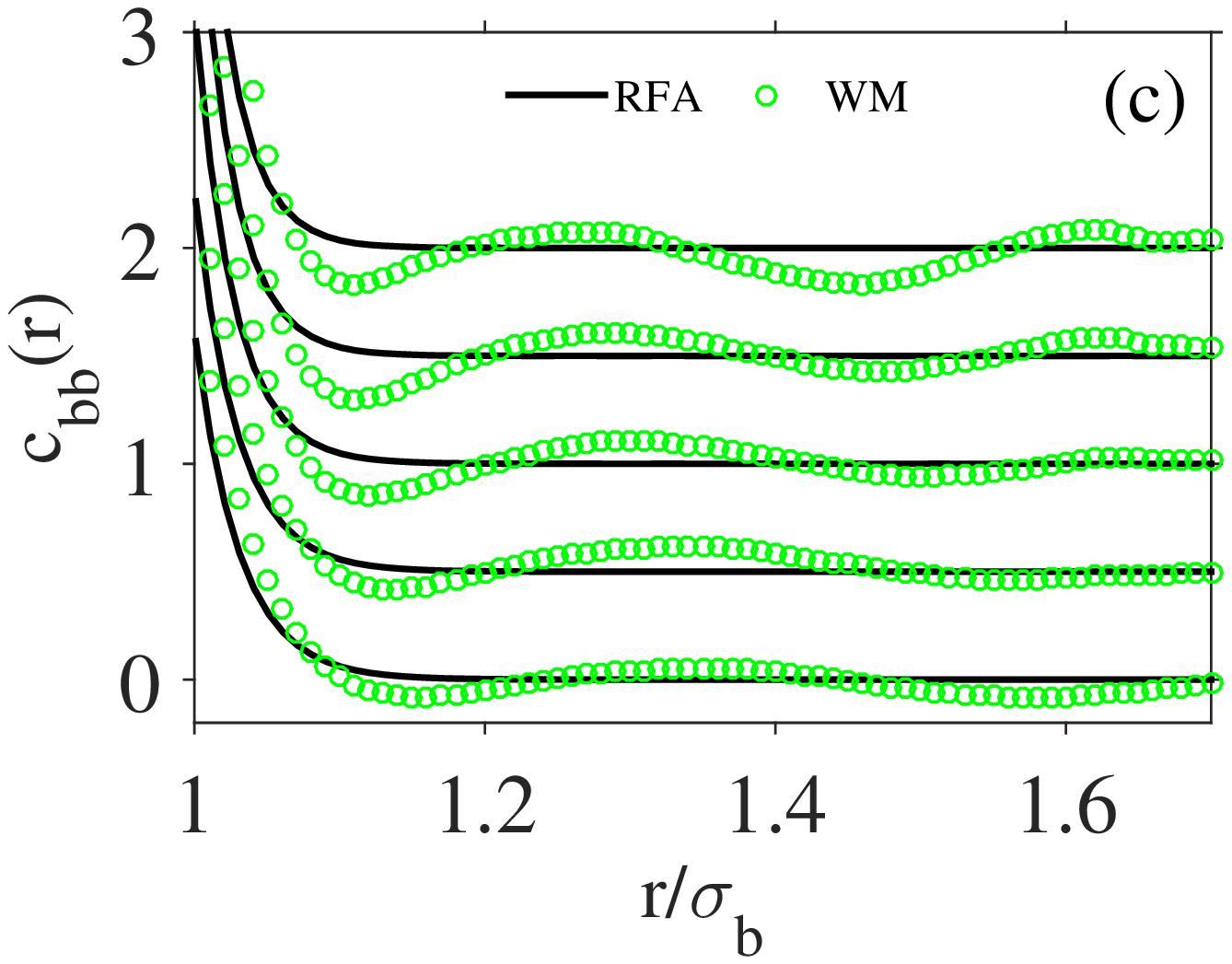}
\caption{Plots of (a) small-small, (b) small-big, and (c) big-big DCFs in the region $r > \q_{ij} $ for additive BHS mixtures with a size ratio $\q_s/\q_b=0.648$ and a  total packing fraction $\eta=0.5$. The partial packing fractions are, from bottom to top, $\eta_s =  0.05, 0.10, 0.20, 0.30, 0.35$. The curves
have been shifted vertically for better clarity.
The open circles represent the results obtained from the WM-scheme and  the solid lines are from the RFA. Note that $c_{ij}^\PY(r)=0$ for $r > \q_{ij} $.}
\label{fig4}
\end{figure*}

Equations (\ref{Eq:hrWM})--(\ref{Eq:hqW&M}), along with the scheme discussed in Sec.\ \ref{sec2}, can be used as a practical means for determining the SCFs (in particular, the DCFs) of additive BHS mixtures.
More explicitly, Eq.\ \eqref{Eq:hqWM} is inserted into Eqs.\ \eqref{Eq:cq11-Eq:cq12} to obtain an analytic form for $\tilde{c}^{WM}_{ij}(k)$ and hence $c_{ij}^{\text{num}}(r)$ by the numerical integration defined by Eq.\ \eqref{Eq:crN}. Also, by expanding $\tilde{h}_{ij}^{WM}(k)$ for large wavenumbers, we obtain a tail form with the structure of Eq.\ \eqref{Eq:hg_tailij} and hence a tail DCF, $\tilde{c}_{ij}^{\text{tail}}(k)$, with the structure of Eqs.\ \eqref{Eq:cqtail11rozw-Eq:cqtail21rozw}; analytical integration then yields $c_{ij}^{\text{tail}}(r)$ from Eq.\ \eqref{Eq:crA}.
In what follows, as done in the monocomponent HS case, we will refer to this as the WM-scheme.

The set of parameters in Eqs.\ (\ref{Eq:hrW&M}) was determined by a nonlinear fitting procedure based on the minimization  of
$\left\vert h_{ij}^{WM}(r)-h_{ij}^{\MD}(r) \right\vert < 10^{-3}$ for each $r/\q_{ij} \in (1,r_c^*)$, where $h_{ij}^{\MD}(r)$ was obtained from our MD simulations. A choice $r_c^* = 5$ was seen to be sufficient for our calculations.

The computation of $h_{ij}^{\MD}(r)$ was performed with the DYNAMO program \cite{BSL11}, for the total packing fraction set to $\eta=0.5$ and partial  packing fractions
$\eta_s = 0.05, 0.1, 0.15, 0.20, 0.22, 0.24, 0.26, 0.28, 0.30, 0.35, 0.40$; the size ratio was fixed at $\q_s/\q_b=0.648$.
These specific conditions were chosen to compare the results for this BHS system with those that were obtained before by simulation and experiment \cite{SPTER16}.
The data for $h_{ij}^{\MD}(r/\q_{ij} <r_c^*)$ must be obtained from long simulations with a large number of particles $(N \sim 10^4)$. Only in this way can the finite-size effects  and the statistical errors in the simulations be reduced sufficiently.
In order to test the $N$-dependence and assess those finite size effects, some calculations were carried out for systems of $N = 2\,916,4\,000,6\,912,8\,788$, and $16\,384$ particles.
It was checked that the simulations for the system of $8\,788$ particles were sufficient to obtain reasonably accurate data.

The histogram grid size of $g_{ij}(r)$ was set to $\delta r/\q_{ij} = 0.01$, which was found to be an optimal choice. The MD simulations were carried out typically for a total number of $2 \times 10^9$ collisions, and the statistical uncertainty of the $h_{ij}^{\MD}(r)$ function was obtained with the block averaging method \cite{AT17}. For each density, and in the whole range  $r/\q_{ij} \in (1,r_c^*)$, the accuracy of $h_{ij}^{\MD}(r)$ was such that the estimated uncertainty was $ < 10^{-3}$, being up to $0.002$ near contact for the highest densities and becoming less than $0.0001$ at larger particle separations. For large systems, the finite-size effects in the MD calculations of the RDF arise mainly from fixing the particle number, i.e., from the relation between canonical and grand-canonical ensembles.
The corrections required to convert data from the MD simulations to the canonical ensemble are of $\mathcal{O}(1/N^2)$ \cite{SDE96,BH01}, which are negligible here. Also, it was checked for
a few densities that the remaining part of the correction factor involving density derivatives was smaller than the obtained data accuracy and therefore could be neglected.

The resulting DCFs, $c_{ij}^{WM}(r)$, are shown in Fig.\ \ref{fig3}, together with the RFA and PY results, in the region $0 < r <\q_{ij}$.
For all three DCFs and all studied partial packing fractions $\eta_s$ there is very good agreement between the WM-scheme and the RFA results.  The agreement with the PY approximation is less satisfactory and deteriorates significantly with increasing packing fraction.
Figure \ref{fig3} is supplemented by Fig.\ \ref{fig4}, where the DCFs are shown in the region $r > \q_{ij}$. As in the monocomponent case [see Fig.\ \ref{fig1}(b)], the WM-scheme shows a (damped) oscillatory behavior, a feature not captured by the RFA.
These small deviations between the monotonic (RFA) and the oscillatory (WM-scheme) results are expected to be reflected in the subleading pole density dependence. Also, it is worth  noticing that the key region $r\gtrsim \q_{ij}$ is very well described by the RFA.

\subsubsection{Determination of the poles of $\tilde{h}_{ij}(k)$}
\begin{figure}[h!]
	\includegraphics[width=0.9\columnwidth]{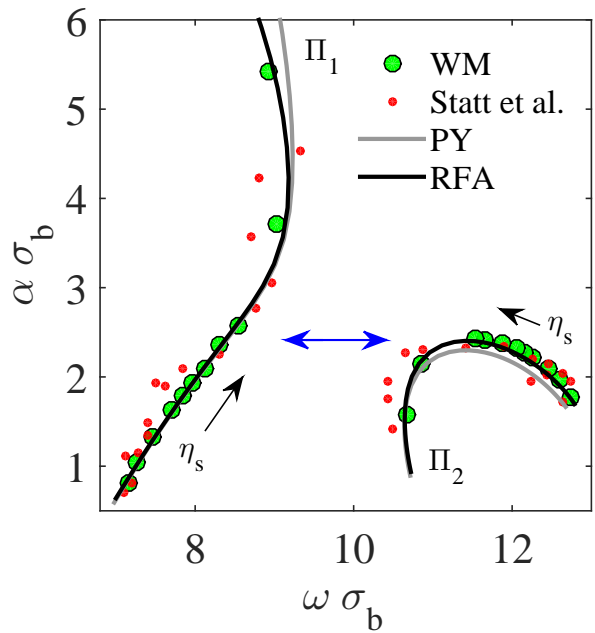}
	\caption{Structure of the two leading poles for additive BHS mixtures  with a size ratio $\q_s/\q_b=0.648$ and a  total packing fraction $\eta=0.5$. Here, $\alpha$ and $\omega$ denote the damping coefficient and the oscillation frequency, respectively (see Eq.\ \eqref{Eq:hr_reduced}). Big (green) circles are results from the WM-scheme (for each $\eta_s$, $\alpha$ and $\omega$ were obtained by solving Eqs.\ \eqref{Eq:poles1-Eq:poles2}), small (red) circles are simulations results from Statt et al.\ \cite{SPTER16}, black solid lines are  from the RFA, and grey solid lines are from the PY approximation. The horizontal blue double-sided arrow indicates the location of the structural crossover, which takes place near $\eta_s=0.29$ and $\talpha \q_b= 2.4$. The left and right poles are denoted as $\Pi_1$ (first) and $\Pi_2$ (second), respectively (see main text).}
	\label{fig5}
\end{figure}

The obtained DCFs, $c_{ij}({r})$, allow for the  determination of the leading poles of $\tilde{h}_{ij}(k)$ by application of the relations in Eqs.\ (\ref{Eq:poles1-Eq:poles2}) and  the results for the first two poles ($\Pi_1$ and $\Pi_2$) are presented in Fig.\ \ref{fig5}.
The pole $\Pi_1$ has $\omega_1\q_b\gtrsim 2\pi$, which corresponds to a wavelength in the oscillatory decay of $h_ {ij}(r)$ comparable to the diameter of the big spheres, while the pole $\Pi_2$ has $\omega_2\q_b\gtrsim 2\pi \q_b/\q_s$, corresponding to a wavelength comparable to the diameter of the small spheres.
An interesting structural crossover \cite{ELHH94,SPTER16} occurs at $\eta_s\simeq 0.29$, such that the leading pole (i.e., the pole with a smaller value of $\alpha$) changes from $\Pi_1$ if $\eta_s<0.29$ to $\Pi_2$ if $\eta_s>0.29$.

The  agreement between the WM-scheme and the theoretical predictions is very  good, especially in the case of the RFA. In fact, the improvement of the RFA over the PY approximation is quite apparent for the subleading pole (i.e., $\Pi_1$ if $\eta_s>0.29$ and  $\Pi_2$ if $\eta_s<0.29$). This subleading pole reflects  the role of the DCFs in the region $r > \sigma_{ij}$ and their  subtle but important influence on the long-range structure of additive BHS fluid mixtures.
We have checked that  the behaviors $c_{ij}(r>\q_{ij})$ do indeed have an impact on the subleading pole by considering `hybrid' DCFs given by the WM-scheme for $r<\q_{ij}$ and either  $c_{ij}(r)=c^\PY_{ij}(r)=0$ or  $c_{ij}(r)=c^\RFA_{ij}(r)$ for $r>\q_{ij}$.
From the results for the monocomponent HS fluid  (see Fig.\ \ref{fig2}), it  may be expected that the performance of the RFA for the subleading pole  deteriorates at higher values of the total packing fraction  of the system. Further studies at other conditions and for different BHS systems are needed to be performed.

In Fig.\ \ref{fig5},  the  recent results by Statt et al.\ \cite{SPTER16}, obtained from a direct fitting of  the RDF simulation data, are also shown. The authors also performed particle resolved experiments on HS like colloids. In general,  the  direct fitting of the $h_{ij}(r)$ functions in a finite domain may  provide ambiguous  information on the leading poles because of possible errors due to factors such as the choice of the distance interval, the number of poles considered in the fitting, and the separation between the poles. However, it is interesting to observe that the results of Ref.\ \onlinecite{SPTER16}  reflect quite well the trends of $\Pi_1$ and $\Pi_2$.

\begin{figure}[b]
	\includegraphics[width=0.9\columnwidth]{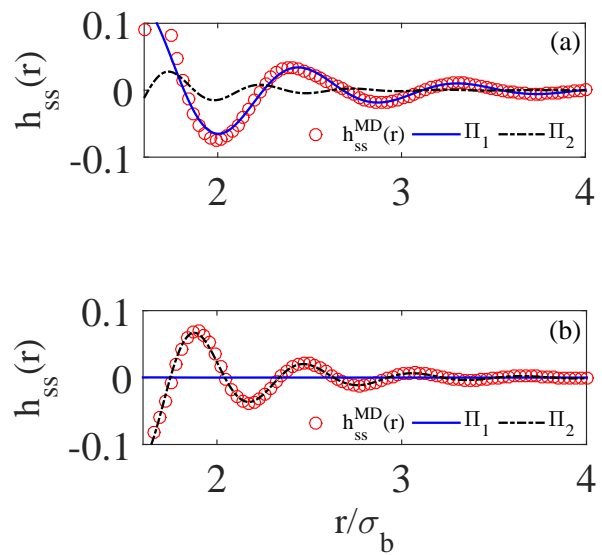}
	\caption{Small-small correlation function $h_{ss}({r})$ for additive BHS mixtures with $\q_s/\q_b=0.648$, $\eta=0.5$, and (a) $\eta_s=0.1$ and (b) $\eta_s=0.4$.  The open (red) circles are the MD data, while the solid (blue) and dot-dashed (black) lines represent the contribution of $\Pi_1$ and $\Pi_2$ pole, respectively. Note a change in the type of the leading pole  describing the decay of the function at $\eta_s =0.1$ ($\Pi_1$) and $\eta_s=0.4$ ($\Pi_2$).}
	\label{fig6}
\end{figure}

The structural crossover phenomenon  is illustrated in Fig.\ \ref{fig6}, where the decay of  $h_{ss}({r})$  at  $\eta_s =0.1$ and $\eta_s=0.4$ is shown.
The amplitudes and phases were calculated from the scheme described in Sec.\ \ref{sec2d}.
It was also verified numerically that the amplitude and phase relations \cite{ELHH94} $A_{ss}A_{bb}=A_{sb}^2$ and $\delta_{ss}+\delta_{bb}=2 \delta_{sb}$ were satisfied.
It is observed that the wavelength of the oscillatory decay is close to $\q_b$ at $\eta_s=0.1$, while it is close to $\q_s$ at $\eta_s=0.4$.
It is also noteworthy that a leading-pole representation of $h_{ss}(r)$ is already quite good, even for not large distances ($r\approx 2\q_b$). A  two-pole representation ($M=2$), including the leading and subleading terms, turns out to be excellent for distances beyond the second maximum.

\section{Conclusions}
\label{sec4}
In this work we have investigated the SCFs of additive BHS  mixtures.
A scheme combining accurate MD simulation data, the pole structure representation of the total correlation functions $h_{ij}(r)$, and the OZ equation has been developed.
It is a nontrivial extension of the approach exploited previously for the monocomponent HS fluid \cite{PBH17}.

An important feature of the presented scheme is that some of the calculations can be performed analytically by taking into account the analytical forms for long  distances (real space) and wave-numbers (Fourier space).
In this way, the DCFs can be determined with great accuracy, thus allowing for calculating  the density dependence of the leading poles and hence the decay of the pair correlation functions $g_{ij}(r)$ of the bulk liquid mixtures.

The obtained results were compared with analytical predictions of the PY and RFA approximations.
In the range of the studied densities,  a very good agreement between  the RFA  and the calculated DCFs is found for separations less than $\sigma_{ij}$. Such  an agreement is observed also for the first pole. In the case of  the second pole for monocomponent fluids, a slight discrepancy for higher densities  is supposed to be caused by an oscillatory form of the DCFs at separations greater than the sphere diameter. In the case of the BHS mixtures analyzed ($\eta=0.5$), however, the agreement is found to be good also for the second pole.
Thus,  our results  indicate that the RFA can predict well the long-range behavior of the SCFs of  BHS mixtures, including the structural crossover, thus representing an improvement over the PY approximation.

\begin{acknowledgments}
S.P. is grateful to the  Universidad de Extremadura, where most of this work was carried out during his scientific internship, which was supported by Grant No. DEC-2018/02/X/ST3/03122 financed by the National Science Center, Poland.
M.L.H. also wishes to thank the hospitality of  Universidad de Extremadura during a short summer visit in which part of this work was done.
S.B.Y and A.S. acknowledge support by the Spanish  Agencia Estatal de Investigaci\'on Grant (partially financed by the ERDF) No.~FIS2016-76359-P and by the
  Junta de Extremadura (Spain) Grant (also partially financed by the ERDF) No.~GR18079.
Some of the calculations were performed at the Pozna\'{n} Supercomputing and Networking Center (PCSS).
\end{acknowledgments}

\appendix
\begin{widetext}
\section{Coefficients in the large-$k$ limit of $\tilde{h}_{ij}(\tk)$}
\label{appA}
In this Appendix, a derivation of the large wavenumber $k$ expressions in Eqs.\ (\ref{Eq:hg_tailij})--(\ref{Eq:hg_tail_C1-D2}) is presented. By making use of the mathematical identities $\sin(a \pm x) = \sin(a)\cos(x) \pm \cos(a) \sin(x)$, $\cos(a \pm x) = \cos(a)\cos(x) \mp \sin(a) \sin(x)$, and $(1 + x)^{-1} = 1 - x +
x^2 - x^3 + \cdots$, it is possible to obtain Eq.\ (\ref{Eq:hg_tailij}) from Eq.\ (\ref{Eq:hq_ij}), the first few coefficients being
\begin{subequations}
\label{APPAEq:hq_tail_C}
\begin{eqnarray}
C_{ij}^{(1)} &=& 4\pi  \q_{ij} +  4\pi  \sum_{n=1}^{\infty} \tA_{ij}^{(n)} e^{-\talpha_{n} \q_{ij}} \sin(\tomega_{n} \q_{ij} + \tdelta_{ij}^{(n)}), \\
D_{ij}^{(1)} &=& -4\pi -4\pi \sum_{n=1}^{\infty} \tA_{ij}^{(n)} e^{-\talpha_{n} \q_{ij}}\left[\tomega_{n} \cos(\tomega_{n} \q_{ij} + \tdelta_{ij}^{(n)})
-\talpha_{n} \sin(\tomega_{n} \q_{ij} + \tdelta_{ij}^{(n)})\right] , \\
C_{ij}^{(2)} &=& -4\pi \sum_{n=1}^{\infty} \tA_{ij}^{(n)} e^{-\talpha_{n} \q_{ij}} \left[(\talpha_{n}^2-\tomega_{n}^2)\sin(\tomega_{n} \q_{ij} +\tdelta_{ij}^{(n)})
 -  2 \talpha_{n} \tomega_{n} \cos(\tomega_{n} \q_{ij} + \tdelta_{ij}^{(n)}) \right],  \\
D_{ij}^{(2)} &=& -4\pi \sum_{n=1}^{\infty} \tA_{ij}^{(n)} e^{-\talpha_{n} \q_{ij}} \left[(\talpha_{n}^3-3\talpha_{n}\tomega_{n}^2)\sin(\tomega_{n} \q_{ij} +\tdelta_{ij}^{(n)})
+(\tomega_{n}^3-3\talpha_{n}^2\tomega_{n}) \cos(\tomega_{n} \q_{ij} +\tdelta_{ij}^{(n)}) \right].
\end{eqnarray}
\end{subequations}

Next, taking derivatives in Eq.\ \eqref{Eq:hr_reduced} (for $\tr>\q_{ij}$), one finds
\begin{subequations}
\label{APPAEq:d1g-APPAEq:d3g}
\begin{eqnarray}
\label{APPAEq:d1g}
{g}_{ij}'(r) &=& \frac{1-{g}_{ij}(r)}{\tr} +  \sum_{n=1}^{\infty} \frac{\tA_{ij}^{(n)}}{\tr} e^{-\talpha_{n} \tr} \left[ \tomega_{n} \cos(\tomega_{n} \tr
+ \tdelta_{ij}^{(n)})  -\talpha_{n} \sin(\tomega_{n} \tr + \tdelta_{ij}^{(n)}) \right],
\\
\label{APPAEq:d2g}
{g}_{ij}''(r) &=& -{\frac{2}{\tr}}{g}_{ij}'(r) + \sum_{n=1}^{\infty} {\frac {\tA_{ij}^{(n)}}{\tr}} e^{-\talpha_{n} \tr} \left[ (\talpha_{n}^2-\tomega_{n}^2)\sin(\tomega_{n} \tr
+\tdelta_{ij}^{(n)})  -2 \talpha_{n} \tomega_{n} \cos(\tomega_{n} \tr + \tdelta_{ij}^{(n)}) \right],
\\
\label{APPAEq:d3g}
{g}_{ij}'''(r) &=& -{\frac{3}{\tr}}{g}_{ij}''(r) -  \sum_{n=1}^{\infty} {\frac {\tA_{ij}^{(n)}}{\tr}} e^{-\talpha_{n} \tr} \left[ (\talpha_{n}^3-3\talpha_{n}\tomega_{n}^2)\sin(\tomega_{n} \tr +\tdelta_{ij}^{(n)})  +(\tomega_{n}^2-3\talpha_{n}^2\tomega_{n}) \cos(\tomega_{n} \tr +\tdelta_{ij}^{(n)}) \right].
\end{eqnarray}
\end{subequations}
Particularizing Eqs.\ \eqref{Eq:hq_ij} and \eqref{APPAEq:d1g-APPAEq:d3g} to $\tr=\q_{ij}^+$, and taking into account Eqs.\ \eqref{APPAEq:hq_tail_C}, it is straightforward to get Eqs.\ \eqref{Eq:hg_tail_C1-D2}.

\section{First few terms in $c^\tail_{ij}(r)$}
\label{appB0}
Let us introduce the mathematical functions
\begin{equation}
\label{I1-I3}
  \mathcal{I}_n(r,a)=\int_Q^\infty dk\, \frac{\sin(k r)\cos(ka)}{k^n},\quad
  \mathcal{J}_n(r,a)=\int_Q^\infty dk\, \frac{\sin(k r)\sin(ka)}{k^n},\quad r,Q>0.
\end{equation}
Thus, insertion of Eqs.\ \eqref{Eq:cqtail11rozw-Eq:cqtail21rozw}  into Eq.\ \eqref{Eq:crA} gives
\begin{subequations}
\label{Eq:cqtailijrozw}
\begin{eqnarray}
c_{ss}^\tail(r)&=&
\frac{C_{ss}^{(1)}}{2\pi^2r} \mathcal{I}_1(r,\q_s) + \frac{D_{ss}^{(1)}}{2\pi^2r}   \mathcal{J}_2(r,\q_s) +
\frac{C_{ss}^{(2)}}{2\pi^2r} \mathcal{I}_3(r,\q_s) - \trho_s  \frac{{C_{ss}^{(1)}}^2}{4\pi^2r}\left[\mathcal{I}_3(r,2\q_s)+\mathcal{I}_3(r,0)\right]\nonumber\\
 &&- \trho_b  \frac{{C_{sb}^{(1)}}^2}{4\pi^2r}\left[\mathcal{I}_3(r,\q_s+\q_b)+\mathcal{I}_3(r,0)\right]+  \frac{D_{ss}^{(2)}}{2\pi^2r} \mathcal{J}_4(r,\q_{s}) - \trho_s  \frac{C_{ss}^{(1)} D_{ss}^{(1)}}{2\pi^2r} \mathcal{J}_4(r,2\q_{s})\nonumber\\
&& - \trho_b  \frac{C_{sb}^{(1)} D_{sb}^{(1)}}{2\pi^2r}\mathcal{J}_4(r,\q_{s}+\q_b)+ \cdots,
\end{eqnarray}
\begin{eqnarray}
c_{sb}^\tail(r)&=&
\frac{C_{sb}^{(1)}}{2\pi^2r}\mathcal{I}_1(r,\q_{sb})  + \frac{D_{sb}^{(1)}}{2\pi^2r} \mathcal{J}_2(r,\q_{sb})
+\frac{C_{sb}^{(2)}}{2\pi^2r}\mathcal{I}_3(r,\q_{sb})  -   \frac{C_{sb}^{(1)}}{4\pi^2r}\left\{\trho_s C_{ss}^{(1)}\left[\mathcal{I}_3(r,\q_{sb}+\q_s)
+\mathcal{I}_3(r,\frac{\q_{b}-\q_s}{2})
\right]\right.
\nonumber \\
&&+\left. \trho_b   C_{bb}^{(1)}\left[\mathcal{I}_3(r,\q_{sb}+\q_b)+\mathcal{I}_3(r,\frac{\q_{b}-\q_s}{2})\right]\right\}
+ \frac{D_{sb}^{(2)}}{2\pi^2r}\mathcal{J}_4(r,\q_{sb})
\nonumber\\
&&
- \frac{D_{sb}^{(1)}   }{4\pi^2r}\left\{ \trho_s C_{ss}^{(1)}\left[\mathcal{J}_4(r,\q_{sb}+\q_s)+\mathcal{J}_4(r,\frac{\q_{b}-\q_s}{2})\right]+  \trho_b C_{bb}^{(1)}\left[\mathcal{J}_4(r,\q_{sb}+\q_b)-\mathcal{J}_4(r,\frac{\q_{b}-\q_s}{2})\right]\right\}  \nonumber \\
&&
-    \frac{C_{sb}^{(1)}    }{4\pi^2r}\left\{\trho_s D_{ss}^{(1)}\left[\mathcal{J}_4(r,\q_{sb}+\q_s)-\mathcal{J}_4(r,\frac{\q_{b}-\q_s}{2})\right]+ \trho_b  D_{bb}^{(1)}\left[\mathcal{J}_4(r,\q_{sb}+\q_b)+\mathcal{J}_4(r,\frac{\q_{b}-\q_s}{2})\right] \right\}+\cdots.\nonumber\\
&&
\end{eqnarray}
\end{subequations}
The exact expression for the function $\mathcal{I}_1$ is
\begin{equation}
	\label{B2}
  \mathcal{I}_1(r,a)=\frac{\pi}{2}\Theta(r-|a|)-\frac{1}{2}\text{Si}(Q(r+a))-
  \frac{1}{2}\text{Si}(Q(r-a)),
\end{equation}
where
\begin{equation}
  \text{Si}(z)\equiv \int_0^z dt\,\frac{\sin t}{t}
  \end{equation}
is the sine integral function.

We observe from Eq.\ \eqref{B2} that $\mathcal{I}_1(r,a)$ presents a discontinuity at $r=a$ of zeroth order. More specifically,
 \begin{equation}
 \label{disc_I1}
\lim_{r\to a^+}\mathcal{I}_1(r,a)-\lim_{r\to a^-}\mathcal{I}_1(r,a)=\frac{\pi}{2}.
\end{equation}
On the other hand, the discontinuities of $\mathcal{J}_2(r,a)$, $\mathcal{I}_3(r,a)$, and $\mathcal{J}_4(r,a)$ at $r=a$ are of first, second, and
third order, respectively. Taking into account Eq.\ \eqref{disc_I1}, it is possible to obtain from Eqs.\ \eqref{Eq:cqtailijrozw} the results displayed in Eqs.\ \eqref{disc_cij_contact}.

\section{Integrals $I_{i}^{(n)}$ and $I_{sb}^{(n)}$}
\label{appB}
In this Appendix the integrals required in Eqs.\ (\ref{Eq:poles1-Eq:poles2}) are presented. They are
\begin{subequations}
\begin{equation}
\label{Eq:int01}
I_{i}^{(0)} = { 4 \pi \over {\tomega}}J_{ii} ,\quad
I_{sb}^{(0)}=-{ 16 \pi^2\over {\talpha^2 + \tomega^2}}\left[J_{ss}J_{bb}-K_{ss}K_{bb}-J_{sb}^2+K_{sb}^2+\frac{\alpha}{\omega}\left(J_{ss}K_{bb}+J_{bb}K_{ss}-2J_{sb}K_{sb}\right)\right],
\end{equation}
\begin{equation}
\label{Eq:int11}
I_{i}^{(1)}= { 4 \pi \over {\talpha}} K_{ii},
\quad
I_{sb}^{(1)}={ 16 \pi^2\over {\talpha^2 + \tomega^2}}\left[J_{ss}J_{bb}-K_{ss}K_{bb}-J_{sb}^2+K_{sb}^2-\frac{\omega}{\alpha}\left(J_{ss}K_{bb}+J_{bb}K_{ss}-2J_{sb}K_{sb}\right)\right]
,
\end{equation}
\end{subequations}
where
\begin{equation}
J_{ij}\equiv\int_0^{\infty}d\tr\, c_{ij}({r}) \tr \cosh(\talpha \tr) \sin(\tomega \tr),\quad
K_{ij}\equiv\int_0^{\infty}d\tr\, c_{ij}({r}) \tr \sinh(\talpha \tr) \cos(\tomega \tr).
\end{equation}
\end{widetext}

\bibliography{BHSmixture}

\end{document}